\documentclass[10pt, a4paper]{article}
\usepackage{lrec2022} % this is the new LREC2022 Style
\usepackage{multibib}
\newcites{languageresource}{Language Resources}
\usepackage{graphicx}
\usepackage{tabularx}
\usepackage{soul}
\usepackage{titlesec}
\titleformat{\section}{\normalfont\large\bfseries\center}{\thesection.}{1em}{}
\titleformat{\subsection}{\normalfont\SmallTitleFont\bfseries\raggedright}{\thesubsection.}{1em}{}
\titleformat{\subsubsection}{\normalfont\normalsize\bfseries\raggedright}{\thesubsubsection.}{1em}{}
\renewcommand\thesection{\arabic{section}}
\renewcommand\thesubsection{\thesection.\arabic{subsection}}
\renewcommand\thesubsubsection{\thesubsection.\arabic{subsubsection}}
\usepackage{epstopdf}
\usepackage[utf8]{inputenc}

\usepackage{hyperref}
\usepackage{xstring}

\usepackage{color}

\usepackage{tikz}
\usetikzlibrary{shapes,arrows}

\title{
A Semi-Automatic Approach to Create Large Gender- and Age-Balanced Speaker Corpora: Usefulness of Speaker Diarization \& Identification. 
}

\name{ Rémi Uro\textsuperscript{1,2}, 
        David Doukhan\textsuperscript{1}, 
        Albert Rilliard\textsuperscript{2,3},
        Laëtitia Larcher\textsuperscript{1},\\
        {\bf \large 
        Anissa-Claire Adgharouamane\textsuperscript{1},
        Marie Tahon\textsuperscript{4}, 
        Antoine Laurent\textsuperscript{4}} } 

\address{\textsuperscript{1}French National Institute of Audiovisual (INA), Paris, France.
    \textsuperscript{2}Université Paris Saclay, CNRS, LISN. \\
    \textsuperscript{3}Universidade Federal do Rio de Janeiro, Brasil.
    \textsuperscript{4}LIUM, Le Mans Université, France. \\
         \{ruro,ddoukhan,llarcher,aadgharouamane\}@ina.fr\\
         albert.rilliard@lisn.fr,    \{marie.tahon,antoine.laurent\}@univ-lemans.fr\\}

\abstract{% 150 to 200 words 
This paper presents a semi-automatic approach to create a diachronic corpus of voices balanced for speaker's age, gender, and recording period, according to 32 categories (2 genders, 4 age ranges and 4 recording periods). 
Corpora were selected at French National Institute of Audiovisual (INA) to obtain at least 30 speakers per category (a total of 960 speakers; only 874 have be found yet).
For each speaker, speech excerpts were extracted from audiovisual documents using an automatic pipeline consisting of speech detection, background music and overlapped speech removal and speaker diarization, used to present clean speaker segments to human annotators identifying target speakers.
This pipeline proved highly effective, cutting down manual processing by a factor of ten.
Evaluation of the quality of the automatic processing and of the final output is provided.
It shows the automatic processing compare to up-to-date process, and that the output provides high quality speech for most of the selected excerpts. % reformuler
This method shows promise for creating large corpora of known target speakers.
 \\ \newline \Keywords{semi-automatic processing, corpus creation, diarization, speaker identification, gender-balanced, age-balanced, speaker corpus, diachrony} }

\begin{document}

\maketitleabstract

\section{Introduction}

%% STATEMENT OF NEED

This paper describes a semi-automatic method for speeding-up speaker corpora construction.
There is a growing need for reference speech corpora featuring reliable and known speaker characteristics. 
Manual annotations of high level features such as speaker segmentation is a time consuming process  \cite{broux2018computer}. 
%We propose to use up-to-date diarization and speaker identification capabilities so to reduce human intervention, and enable the creation of large spoken corpora at a minimum cost.
We propose to use up-to-date diarization and speaker identification to reduce human intervention, and enable the creation of large spoken corpora at a minimum cost.

This work is part of the Gender Equality Monitor (GEM) project, %that aims at enabling the use of Natural Language Processing tools for the analysis of potential gender bias in broadcast media in France.
that aims at describing male and female representation differences in the French broadcast media at scale, using automatic information extraction methods.
A main institution involved in the project is the French National Institute of Audiovisual (INA), a public institution in charge of archiving French audiovisual heritage (so called legal deposit). INA's collections include 23 million hours of TV and radio programs broadcasted since the 1930's.% , which has as one of its main0 attribution the archiving and enhancement of media broadcasted in France (so called legal deposit). 
%Its large archives include systematic recordings of some categories of audiovisual media since the 1940's.
% A AMELIORER

First, the targeted characteristics of the corpus are presented, with a discussion of comparable datasets, and the potential use of this resource.
A literature review on diarization and speaker identification systems is then presented.
The rest of the paper presents: (1) the methodological steps for the creation of this corpus, (2) evaluations of these steps in terms of performance on available data with comparable characteristics, (3) a subjective evaluation of the quality of voice of the final output, and (4) an estimation of processing time reduction compared to manual processing.

\section{Targeted corpus and existing resources}

\subsection{Voices \& gender over time}

For the needs of the GEM project, the capacity to describe the acoustic characteristics of a given speaker's voice is important, to compare to its socially recognized characteristics, that predominantly includes the speaker's gender and age.
Having a reliable statistical representation (i.e., based on representative data for several age and gender categories) of the acoustic variation of voices presented in the public arena allows sociological analysis of gender representation, gender stereotypical characteristics, potential changes in voice as a device to present one's self public image, etc.
The voice is an important aspect of the construction of individual social persona or character \cite{podesva_phonation_2007,sadanobu_characters_2015}; it changes according to our role in society –- e.g., while talking as an employee to a superior, as a professor to students, as a parent to children, as a friend to sporting mates etc. 
Our voice is developed during childhood and adolescence, varying because of our physiological development \cite{vorperian_developmental_2011}, but also as a culturally shaped representation of our social projection as an individual \cite{sergeant_gender_2009,guzman_acoustic_2014,scott_neural_2022}.
Gender is a key aspect of the construction of voices, and is culturally shaped: speakers have been shown to have different mean pitch in different cultures, and to change their pitch for reaching culture-matching expectations, notably on gender \cite{van_bezooijen_sociocultural_1995,ohara_gender-dependent_1992,ohara_finding_2001}.
Changes in gender-related vocal cues have effects on many aspects of social interaction, pitch having been related to a series of characteristics like credibility, charisma, cuteness, sexual attractiveness, etc. \cite{geiselman_incidental_1983,niebuhr_what_2016,jang_how_2021,gussenhoven_foundations_2016}.

To better document the aspects of self that our voices may voluntarily or not display \cite{scott_neural_2022}, a reference corpus of voices is a great tool, allowing studies on voice presentation in the public arena via broadcast media in France (the GEM project is centered on this country).
%As a culturally build device, voices change according to cultures; the GEM project if centered on French medias, so the limitation to main French broadcasters, if it certainly limits the scope of the corpus socially and cross-culturally.
This corpus would ideally contain voices used in comparable dialogue situations -- i.e., not expressive situations where voice may be especially loud, or express emotions, two aspects which have important effects on vocal characteristics \cite{titze_vocal_1992,lienard_effect_1999,traunmuller_acoustic_2000,lienard_quantifying_2019,goudbeek_beyond_2010}. Although such phenomena may be interesting to describe, they add variability that would require more data to be controlled for.
Thus broadcast programs featuring interviews or discussions with invited people were selected, mostly those recorded in studio situations.
Note that audiovisual program metadata does not allow a full control of the recording settings, especially as it is common to find reporting on a given topic, that may happens to feature the target speaker.
Details on the audiovisual document selection process are given in the next section.

To achieve accurate representation of voices, through their long-term acoustic qualities, it is mandatory to obtain a duration that allows acoustic characteristics to stabilize over articulation and other dynamic aspects. 
\newcite{lofqvist_long-time_1987} showed Long Term Average Spectrum is stabilized at about 20 seconds of continuously voiced speech -- so about twice this duration for raw speech; see also \newcite{arantes_temporal_2014} on prosody. 
We set a lower limit of three minutes of diarized speech per speaker.
It was thought important so to limit the influence of noises and potential backchannels.

\subsection{Available resources}
%reco locuteur \& diarization\\
%robustesse de la reco en fonction de l'age \\
%reconnaissance HF \& etudes diachroniques \cite{doukhan2018describing}.\\
%gender \& age biases See mahault garnerin\\
%diachronic prosodic analysis\\

As far as we know, there is no available speech corpora providing balanced distribution of speaker gender, age and recording date for French media.
%Obtaining such data from existing corpus is not straightforward.
%There are no corpus that we know about that gives information on the speaker's gender and age (with good representation of both sets of categories), together with a diachronic reproduction of such gender and age distribution of speakers.

% corpus with gender information:
%Many corpus present gender information.
%The multilingual corpus described in \citelanguageresource{muthusamy92_icslp} propose gender information, but mostly present prepared speech, and do not have diachronic aspect.
%This is the case of many of the available speech corpus designed for NLP training.
%In the corpus used for \citelanguageresource{yaguchi_speech_2010} study on American English in professional interactions, thee speaker's gender is specified.
%The Switchboard corpus \citelanguageresource{godfrey_switchboard_1992} presents age and gender data on a large amount of American English speakers uttering elicited speech.
%Similar in design, but larger, more recent corpus exist -- e.g. \cite{guo_didispeech_2021} in Mandarin read speech, or the LibriSpeech database for English \citelanguageresource{panayotov_librispeech_2015}.

For resources in French, various corpora are available via the cocoon website\footnote{\url{https://cocoon.huma-num.fr/exist/crdo}} -- that mostly proposes dialectological, sociological or ethnological resources. 
The quality, type of speech, and information available about each speaker is highly variable.
One of these resources, the ESLO corpus \citelanguageresource{baude_elso}, proposes interviews of many residents of Orleans city, made at two periods: beginning of 1970's and of 2010's \cite{eshkol-taravella_grand_2011}. 
If an interesting resource, it is strictly restricted to one city and two time periods; it is also based on field recordings of variable recording qualities.

Another set of corpora was developed during evaluation campaigns for NLP tools. 
For example, the ESTER corpus provides sounds and transcriptions for one hundred hours of broadcast news from French media \citelanguageresource{galliano_ester_2009}.
The ETAPE corpus is a follow-on development with less prepared speech from other types of radio or TV programs \citelanguageresource{gravier_etape_2012}.
These corpora contain information on speaker gender, but not directly on speaker age.
Unfortunately, none of them present a diachronic dimension.
One resource, the Eurodelphes database \cite{barras_transcribing_2002}, proposes a set of broadcast documents spread from the 1940's up to the 1990's; it is nonetheless relatively limited in size for the oldest decades, and is highly unbalanced for gender and age \cite{boula_de_mareuil_diachronic_2012}, thus not suited for the targeted use. 

The corpus presented in \newcite{suire:hal-03097705} is the closest to what we are trying to build.
It is based on media programs with about thirty speakers of each gender selected by period of ten years, over seven decades, with about 5 seconds of speech for each speakers; the age of speakers was not informed. 
These two limitations (short extracts, and no information on speaker's age) make it unsuitable for the study.

%\subsection{NLP tools for corpus processing}
\subsection{Existing (semi-)automatic speaker corpora}

%% Automatisation partielle ou totale
Building audiovisual speaker databases is costly, since it requires finding speakers in audiovisual documents and obtaining the time codes for speakers' speech turns.
To that aim, few fully or partially automated approaches were proposed for building large speaker databases with minimal human involvement.

INA's speaker dictionary was prepared using a semi-automatic procedure based on unsupervised speaker segmentation (diarization) and Optical Character Recognition (OCR) \cite{salmon2014effortless,vallet2016speech}. OCR decoded embedded text presenting people in TV news programs, and filtered characters corresponding to the first twenty thousand most referenced people in INA's audiovisual databases. Speech segments corresponding to decoded embedded text were on a second stage presented to human annotators in charge of stating if speech segments were corresponding to the decoded person name, resulting in an average involvement time of 22,8 seconds per segment.

VoxCeleb was built from YouTube videos using a fully automatic procedure \cite{Nagrani17}. YouTube video queries were applied using target speakers' names and the word `interview`. The set of target speakers was a subset of celebrities known by a preexisting face verification classifier. Active speaker verification models were  used to detect the portions of video with facial lip movements synchronized with the audio track. Face candidates were in a latter stage presented to the face verification classifier using a high threshold.

While these approaches allowed to build large speaker collections, they suffer from several limitations. Both of them require video material, and cannot be used for processing radio collections. INA's speaker dictionary's strategy requires embedded text to be displayed during target's speech turn, while VoxCeleb's strategy requires target speaker's face to be already known by a face verification classifier. Moreover, YouTube's queries do not allow search by speakers' age.

%- Diarization state of the art
%- Speaker identification (state of the art
%- Evaluation of speaker ID ?

The corpus-building process aims at creating a large corpus (more than 960 speakers) of gender- and age- balanced speakers that may serve as a reference of voice qualities (linked to gender representation) presented in broadcast media, from the 1950's until now.
As a manual gathering is out of reach, we aimed at limiting to a minimum manual intervention so to speed the process.
%so we devised the method which is proposed here, that applies diarization and other techniques of natural language processing, so to select segments that contains only speech (i.e., without musical or noise background, or at a low signal-to-noise ratio), so to allow reliable acoustic measurements on the speakers' voices.
This paper present the details of the approach, with an evaluation of the automatized process, and of the quality of the obtained dataset, as well as an estimation of the time gained through the semi-automated method, compared to a fully handmade process.

\section{Methods}

\begin{figure}
    \centering
    \includegraphics[width=0.4\textwidth]{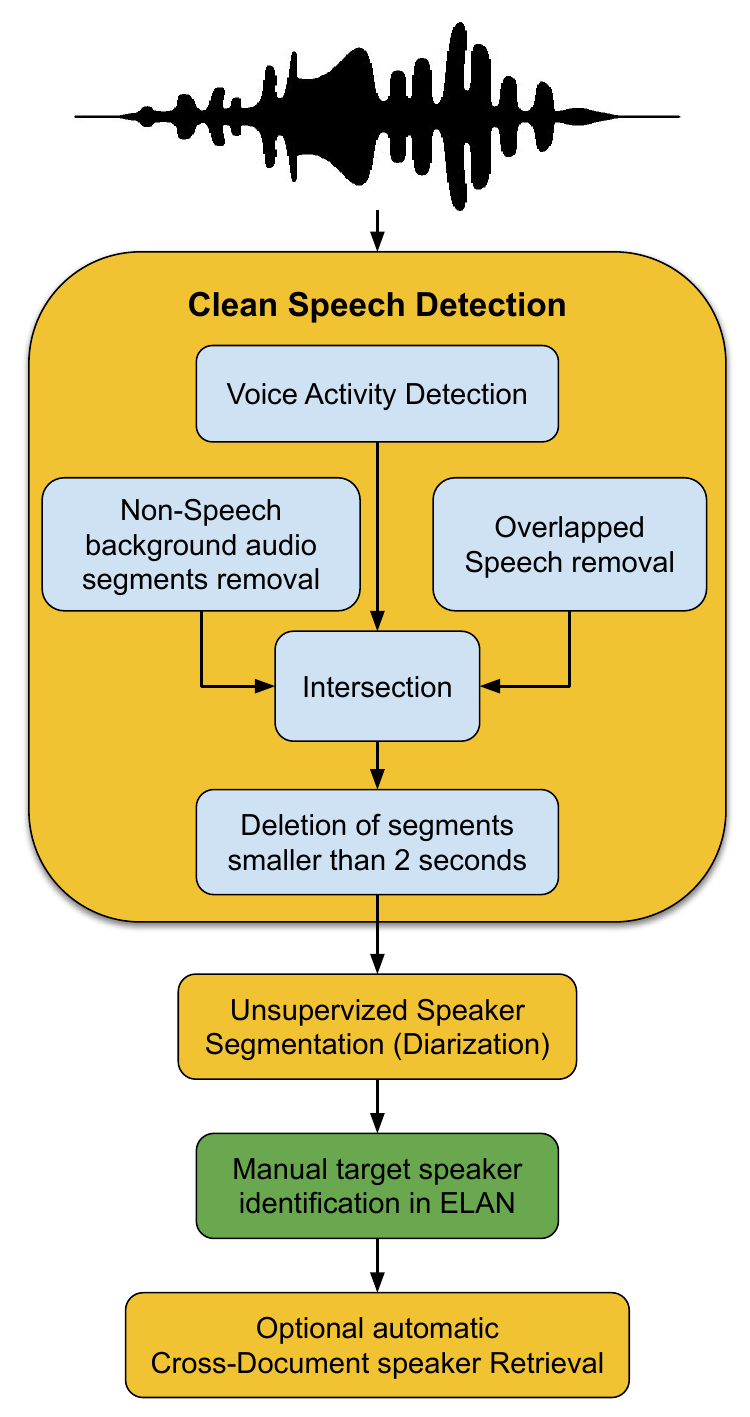}
    \caption{Semi-automatic pipeline proposed for the extraction of \textit{clean} target speaker segments}
    \label{fig:pipeline}
\end{figure}

Speaker selection guidelines were defined to obtain a diachronic speaker corpus with balanced gender, age, and recording period. 
These guidelines were provided to INA's archivists to constitute a corpus of audiovisual documents containing a balanced amount of target speakers.
Figure \ref{fig:pipeline} show the semi-automatic pipeline designed to extract target speaker excerpts from this huge collection of audiovisual documents with a minimal amount of human involvement. 
A \textit{Clean Speech Detection} process was defined to discard speech segments with acoustic properties that may interfere with acoustic parameter extraction. %: speech overlap, high levels of background noise or music.
An unsupervised speaker segmentation and clustering procedure (diarization) was used on the resulting clean speech segments in order to assign numeric identifiers to each speaker found in the recording.
Resulting speaker segmentations were then presented to human annotators in charge of the manual identification of the target speakers.
If the found speaker excerpts in the manually processed documents are less than three minutes,  these excerpts were used to perform automatic cross-document speaker retrieval, to complete the speaker sample in the corpus.

%\subsection{Selection of target speakers}

%\subsection{Diachronic speaker corpus definition}
% Choix effectués (but du corpus => ALBERT)

%As explained above, the resource needs to contain voice samples of a sufficient quality so to extract reliable long term spectral and prosodic data for each speaker -- thus the minimum target of three minutes of voice per speaker.
%To avoid vocal changes linked to stylistic variations, programs featuring dialog and interview are prioritized; this is done because they are also frequently recorded indoor.
%Four age groups were defined, excluding children: 20 to 35, 36 to 50, 51 to 65, and over 65 year old.
%This was based on known changes in voice linked to age and gender \cite{sataloff_chapter_2017,yamauchi_quantitative_2015}.
%Four periods of time were defined, that spans from the 1950's to the 2010's: years 1955-1956, 1975-1976, 1995-1996, 2015-2016.
%This is somewhat arbitrary, but these periods were chosen because it is harder to find archives before the 1950's, especially featuring female speakers, and four periods were though sufficient for our goal, and limit an already large amount of target speakers.

\subsection{Balanced diachronic speaker corpus definition}

%% AR: commenté pour la place (deja dit plus haut)
%The proposed speaker corpus is defined to contain voice samples of a sufficient quality and duration so to extract reliable long term spectral and prosodic features.
%The minimal amount of speech per speaker was set to three minutes, and could either be obtained from a single or from several documents.

To avoid environmental noise and vocal changes linked to stylistic variations, broadcast programs featuring dialog and interviews, recorded indoors, are prioritised.

A total of 30 distinct speakers is required for each of \textbf{32 adult speaker categories} (adding up to a total of at least 960 different speakers), based on the combination of 3 parameters:
\textbf{gender} (male and female), 
\textbf{age} (4 groups: 20 to 35, 35 to 50, 51 to 65 and over 65 years old), and 
\textbf{periods} of time (4 periods: 1955-1956, 1975-1976, 1995-1996, 2015-2016). 
The age groups were based on known changes in voice linked to age and gender \cite{sataloff_chapter_2017,yamauchi_quantitative_2015}.
The periods of time span from the 1950's to the 2010's with 20 years intervals: this is somewhat arbitrary, but these periods were chosen because it is harder to find archives before the 1950's, especially featuring female speakers, and four periods were thought sufficient, and limit the amount of target speakers.
Each speaker was selected only in one of the 32 categories to avoid statistical dependency; we are thinking about preparing a section of the corpus with longitudinal recordings of speakers available in three to four time periods.

\subsection{Audiovisual document selection}
\label{sec:archivist}
% Méthode de selection des cibles (W documentaire)
Document selection based on the corpus definition was realized by INA's archivists, and required 3 weeks of work.
The list of participants and the date of diffusion was extracted from TV and radio archives meta-data, and linked to INA's personality thesaurus to obtain date of birth and gender information. 
This allowed to assign each unique speaker (our "target speakers") to one of the 32 gender, age, and period categories.
A manual selection of 450 TV and radio shows was realized, that usually feature reasonably long studio-recorded interviews, well-known personalities, low amounts of background music and noise, and a low amount of conflicting interactions between participants.

While the number of male found for each category turned out to be greater than thirty, several categories of female had to benefit from additional manual retrieval in INA's collections to reach the minimum amount of speakers required to meet corpus specifications. Female stand out more rarely in broadcast media \cite{doukhan2018describing}, and their number was particularly scarce within the older age groups and in the oldest periods in INA thesaurus, possibly reflecting the gender bias of media presentation and representation \cite{Tuchman2000}.
Research for complementary speakers was necessary and resulted in the enrichment of INA's personality thesaurus, with the inclusion of the characteristics of 182 media personalities.

Table \ref{tab:archivist} presents the characteristics of the corpus constituted by INA's archivists, with 25092 entries for 6051 distinct speakers, distributed between 17307 audiovisual documents, for a total duration of about 393 days of content (9400 hours). 
This is clearly too large to consider a complete manual exploration, and requires the semi-automatic procedure described below.

\begin{table}
    \centering
    \begin{tabular}{|c|c|c|c|c|}
    \hline
    Period & \# TV  & \# Radio  & \#  Unique & Duration \\
           & docs   &   docs    &   speakers & (days)\\
    \hline
    1955-56 & 133 & 508 & 594 & 11\\ 
    1975-76 & 849 & 642 & 1220 & 46\\
    1995-96 & 1565 & 4686 & 2393 & 176 \\
    2015-16 & 933 & 7991 &  1845 & 160\\
    \hline
    \end{tabular}
    \caption{Characteristics of the speaker corpus collected by INA's archivists: number of documents (docs) from TV and radio, of speakers, and total duration.}
    \label{tab:archivist}
\end{table}

% Précisions sur les cibles identifiées
%\subsection{Archives identification}
%??? est ce que ca vaut le coup d'en parler? (david?)
% peut etre pas; pas la place dee toute facon

%\subsection{Diarization \& Speech quality estimation}
% includes manual selection process (not a big thing to describe)
% must contain an estimation of the TIME it took.
%  TIME for setting up the processing algorithms
%  TIME for running the processing 

% WIP 
\subsection{Clean Speech Detection (CSD)}
%\subsection{Cleaning steps}

A \textit{Clean Speech Detection} procedure was proposed to detect the cleanest speech excerpts, suited to prosodic parameter extraction.
Figure~\ref{fig:pipeline} describes the components of our proposal, which allows to obtain speech segments with low amount of overlapped speech, background music or noise.
Speech segments shorter than 2 seconds were rejected, as they are likely to be only small parts of sentences and hence be of little interest to achieve an accurate representation of the voices.

\subsubsection{Voice Activity Detection (VAD) and Overlapped Speech Detection (OVL)}

Voice activity detection was performed using \texttt{InaSpeechSegmenter} \cite{ddoukhanicassp2018}. This system is based on a CNN architecture trained to distinguish speech from music and noise. 
\texttt{InaSpeechSegmenter} was ranked in first position for the VAD task of MIREX 2018 challenge, containing TV and radio corpora which are representative of our target material\footnote{\url{https://www.music-ir.org/mirex/wiki/}}.

%The performances of this system are described in Table \ref{tab:VAD}. 
%Mirex 2018's results were obtained in challenge conditions, using TV and radio corpora, which are representative of our target material. 
%Our proposed system was ranked in first position for the VAD task in this challenge. 
% Worst results were obtained on DIHARD 2 development dataset, known to be harder and contains heterogeneous material (low volume background speech, etc...).
% We target clear speech: undetected low speech is not of interest.

% http://languagelog.ldc.upenn.edu/myl/RyantDIHARD_IS2019.pdf
% https://halshs.archives-ouvertes.fr/halshs-02899402/document

% \begin{table}
%     \centering
%     \begin{tabular}{|c|c|c|c|}
%     \hline
%     Dataset &  precision & recall & f-measure \\
%     \hline
%     Mirex - Dataset 1 & 90.9 & 92.9 & 91.9 \\
%     Mirex - Dataset 2 & 96.0 & 95.6 & 95.8 \\
%     DIHARD 2 Dev & 90.1 & 84.6 & 87.3 \\
%     \hline
%     \end{tabular}
%     \caption{Voice Activity Detection performances on Mirex 2018 (challenge conditions) and on DIHARD 2 development dataset.}
%     \label{tab:VAD}
% \end{table}

% AR: ca faisait une partie tres courte: je l'ai colle a la precedente
%\subsubsection{Overlapped speech detection with pyannote-audio}
In order to isolate the segments corresponding to only one speaker, \texttt{pyannote-audio}'s overlap speech detection \cite{Bredin2020,Bullock2020} was also used. The detected overlapping speech segments were then cut from the initial VAD.

%\subsubsection{Music detection with Spleeter}
\subsubsection{Non-Speech audio event detection}

Audiovisual documents may contain non-speech audio events overlapped with speech, such as music or noise. Such events may interfere with vocal feature estimation and excerpts with these events should be discarded.

A non-speech audio event detection model is proposed, based on \texttt{spleeter} source separation framework ~\cite{hennequin2020spleeter}.
%We used \texttt{spleeter} 2 stems model, designed for separating singing voice from its instrumental accompaniment.
We used \texttt{spleeter} vocals and instrumental accompaniment separation model.
%, and observed good performances for the separation of spoken voice from other audio events (music and noise).
The potential presence of non-speech audio events was linked to the energy of the extracted instrumental accompaniment track, estimated using the root mean square of signal with 200 ms window size and 100 ms hop size. The energy is filtered using a median filter of size 11, and a threshold set at 5\% was used.

The evaluation of the non-speech audio detection model is difficult, due to a lack of annotated resources containing overlapping speech, music and noise annotations.
%While there are several resources suited for speech detection, resources suited for music detection are scarcer, and we do not know annotated audio resources containing overlapping speech, music and noise annotations. 
Table \ref{tab:spleeterdetect} presents the evaluation of our proposal on OpenBMAT, a database of audio streams with annotated music levels \citelanguageresource{melendez_catalan_blai_2019_3381249}. 
With respect to our use-case (obtaining clean speech, even with a low document coverage rate) and to the availability of annotated resources (no dataset with speech, music and noise annotations), we used \texttt{sed\_eval} segment-based detection recall for estimating the performance of our proposal, using time tolerance (collar) of 1 second ~\cite{mesaros2016metrics}. The results show detection performance above 90\% for audible music, and 65\% for hard-to-hear background music, which shall less affect acoustic analysis.

%As we only want clean speech, we also exclude speech overlapped with music or noise. The tool \texttt{spleeter} is used to separate voice and music \cite{hennequin2020spleeter}. If the energy in the resulting music channel is higher then a certain threshold, we consider it as music and exclude those segments from the VAD.

\begin{table}
    \centering
    \begin{tabular}{|c|c|c|c|c|c|}
    \hline
    Level &  bgvl & bg & similar & fg & music \\
    \hline
    Recall (\%)  & 65 & 89.9 & 97.0 & 97.2 & 99 \\ 
    \hline
    \end{tabular}
    \caption{Music detection recall obtained on OpenBMAT for varying music levels: bgvl (hard-to-hear background music), bg (background music), similar (music and other signals mixed at similar levels), fg (foreground music), music (music only)}
    \label{tab:spleeterdetect}
\end{table}

\subsubsection{Clean Speech Detection pipeline coverage}

We tested our pre-processing pipeline on the DIHARD II Development set~\cite{ryant_second_2019}.
Table \ref{tab:preprocess_dur} shows the duration of detected speech at different stages of the pre-processing.%: VAD only; VAD with overlapping speech removed (VAD+OVL); VAD with non-speech events removed (VAD+NSE); VAD with both overlapping speech and non-speech events removed (VAD+OVL+NSE); and full pre-processing pipeline (CSD), i.e. the VAD with overlapping speech and non-speech events removed and segments less than 2 seconds removed. It also shows the duration of speech remaining from the reference after removing overlapping speech (Ref+OVL), and after removing non-speech events (Ref+NSE).
DIHARD II focuses on \textit{hard} diarization, i.e. with lots of low volume background speech, \textit{in the wild} speech  with music, noise and overlapping speech. Note that we target clean speech: low level noisy speech is not of interest for our intended prosodic analyses.
%known to be harder and contains heterogeneous material (low volume background speech, etc...).
% We target clear speech: undetected low speech is not of interest.
Our CSD system eliminates more than half of the total speech time. Since we value a better precision than recall, considering that only about 40\% of a given corpus is usable seems enough. 
Moreover, one can assume that the TV and radio broadcast documents that this system targets should contain more clean speech than DIHARD II documents. 
So the process shall remove a smaller part of the content on such documents.

\begin{table}
    \centering
    \begin{tabular}{|l|r|r|}\hline
         Method & Duration (s) & Coverage  \\\hline
         Reference & 72311 & 100\%\\\hline
         Ref+OVL & 63781 & 88.2\%\\
         Ref+NSE & 32013 & 44.3\%\\\hline
         VAD & 69247 & 89.1\%\\
         VAD+OVL & 56953 & 78.8\%\\
         VAD+NSE & 30166 & 41.7\%\\
         VAD+OVL+NSE & 28360 & 39.2\%\\
         CSD & 23980 & \textbf{33.2\%}\\ 
         \hline
    \end{tabular}
    \caption{Duration of detected speech on the DIHARD II Dev set for the different pre-processing steps and coverage relatively to the reference. (VAD: only VAD; OVL: overlapped speech removal; NSE: non-speech events removal; CSD: clean speech detection --VAD+OVL+NSE + removal of segments less than 2 seconds)}
    \label{tab:preprocess_dur}
\end{table}

\subsection{Diarization with VBx}

We used our CSD as an input VAD for diarization, meaning that \textit{clean} speech is considered as \textit{speech} and \textit{non-clean} speech as \textit{non-speech}.
We use the $x$-vector based diarization system VBx \cite{landini2022bayesian} with the \texttt{ResNet101\_16kHz} model, pretrained on VoxCeleb1 \cite{Nagrani17}, VoxCeleb2 \cite{Chung18b} and CN-CELEB \cite{fan_cnceleb_2020}.
The diarization step outputs clusters id corresponding to a unique speaker.

We have evaluated the VBx model using our CSD on the DIHARD II Development set~\citelanguageresource{ryant_second_2019}.
Table \ref{tab:der_dihard} shows the Diariation Error Rate (DER) for the different stages of pre-processing. The DER is computed by removing \textit{non-clean} segments from the reference.
As expected, we observe a better DER when \textit{non-clean} speech segments are removed, the diarization task being easier. 
We obtained a DER of 14.7 with a 0.25s collar using our pre-processing pipeline, which is comparable to the DER of 12.23\% obtained by \cite{landini2022bayesian} with oracle VAD.
%

%Using the reference VAD, we obtain a DER similar to the DIHARD II baseline performances. 
\begin{table}[]
    \centering
    \begin{tabular}{|l|r|}\hline
         Input & DER (\%) \\\hline
         Reference & 23.8\\
         VAD & 24.5\\
         VAD+OVL & 21.3\\
         VAD+NSE & 16.5\\
         CSD & \textbf{14.7}\\\hline
    \end{tabular}
    \caption{Performance of the diarization system (measured by DER) on DIHARD II dev set for the different stages of pre-processing. (collar=0.25)}
    \label{tab:der_dihard}
\end{table}

%\subsection{Manual within-show speaker identification}
\subsection{Manual speaker identification}
\label{sec:man_id}
Each audiovisual document of the source corpus was associated to a list of target speakers with known age, gender and role (anchor, participant) provided by INA's archivists (see section \ref{sec:archivist}), to be manually identified. %in order to obtain a diachronic age and gender balanced speaker corpus.
The \textit{clean speech diarization} described above was exported to ELAN video annotation tool and presented to human annotators, together with the list of target speakers \cite{sloetjes_annotation_2008}. 

For each document, annotators had to map diarization cluster id's to target speakers' identities.%the annotator has to identify one or multiple speakers according to what is known from INA documentation (name, gender, age etc.). 
%Each identified speaker is associated with their corresponding cluster id. 

%In order to assist the annotator with the task of identifying the speaker, information about their gender, age and role in the document (anchor, participant) are also provided.\\
The complexity of this task varies a lot depending on the type of document and the role of the target. For instance a recent TV interview with only two speakers can be processed in a few seconds, whereas an old radio show with multiple characters, mostly unknown nowadays, may require to listen almost all the document, and sometimes the use of internet to find photos or details to spot the target.

%\subsection{Identification of known speakers}
\subsection{Automatic cross-show speaker identification}
The corpus aims at presenting at least three minutes of speech by speaker. 
For most documents, the manual identification described in \ref{sec:man_id} was enough because the documents were chosen to maximise the speaking time of the target speakers.
However, it is not the case for all documents.
Then, the segments linked to the target speaker in the manually annotated document were used as a reference to automatically identify this speaker in other documents.

Using VBx $x$-vector extraction model, we retrieve one $x$-vector per segment corresponding to the target speaker in the annotated document, giving us a matrix $x_{known}$. 
The non-annotated document was then pre-processed and automatically diarized in the same way, and one $x$-vector per segment of this document was extracted, giving us a matrix $x_{target}$.
Cosine similarity was computed between all the vectors in $x_{known}$ and $x_{target}$. For each vector in $x_{target}$, the mean similarity to the $x_{known}$ vectors gives the probability score of the vector corresponding to the target speaker. If this score exceeds a given threshold, we considered that the segment corresponds to the target speaker. If no segment had a score above the threshold, we considered the target speaker absent from the document.
We chose to focus on segment-level identification in order to maximize the precision, even though it may increase the false negative rate.

\begin{figure}
    \centering
    \includegraphics[width=.5\textwidth]{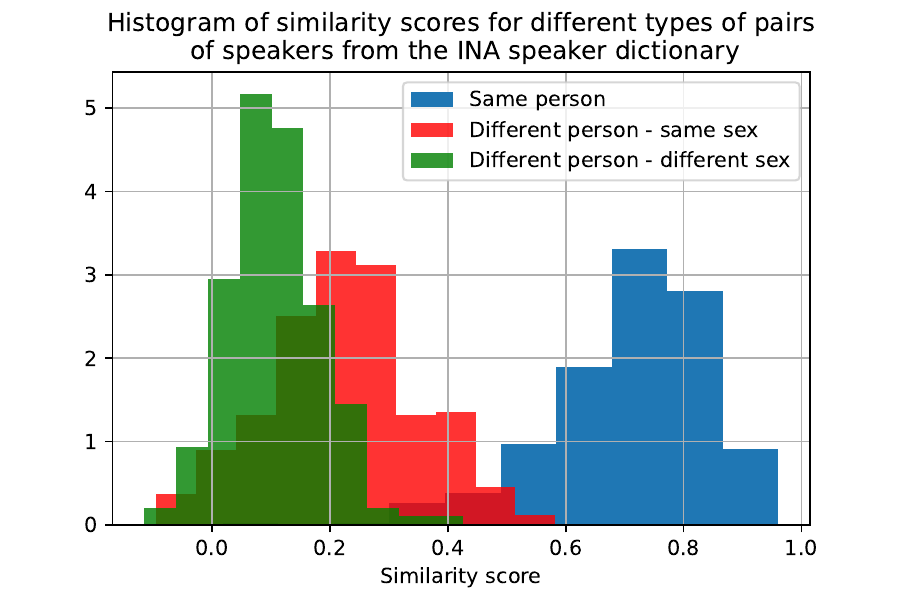}
    \caption{Histogram of similarity scores for different types of pairs of speakers from the INA speaker dictionary}
    \label{fig:hist_sim_scors}
\end{figure}

We evaluated our speaker identification system on INA speaker dictionary~\cite{vallet2016speech} which contains materials extracted from French TV archives and is similar to the contents our system was designed for. This dataset contains about 1300 speakers extracted from French TV broadcasts. We analyse the similarity score between three different types of recording pairs: the same speaker in two different recording sessions, two different speakers of the same gender and two different speakers of different gender.
This was evaluated on a gender-balanced subset of 718 speakers allowing for each speaker to be in at least two recording sessions, ending up with 359 pairs for each type. The mean duration of the used segments was 14s.
Figure \ref{fig:hist_sim_scors} shows the distribution of similarity scores for the different types of pairs. There is few overlap between the scores corresponding to the same speaker and those corresponding to different speakers. There is very little (4\%) overlap between scores for the same speaker and scores for speakers of different genders. 

\begin{figure}
    \centering
    \includegraphics[width=.5\textwidth]{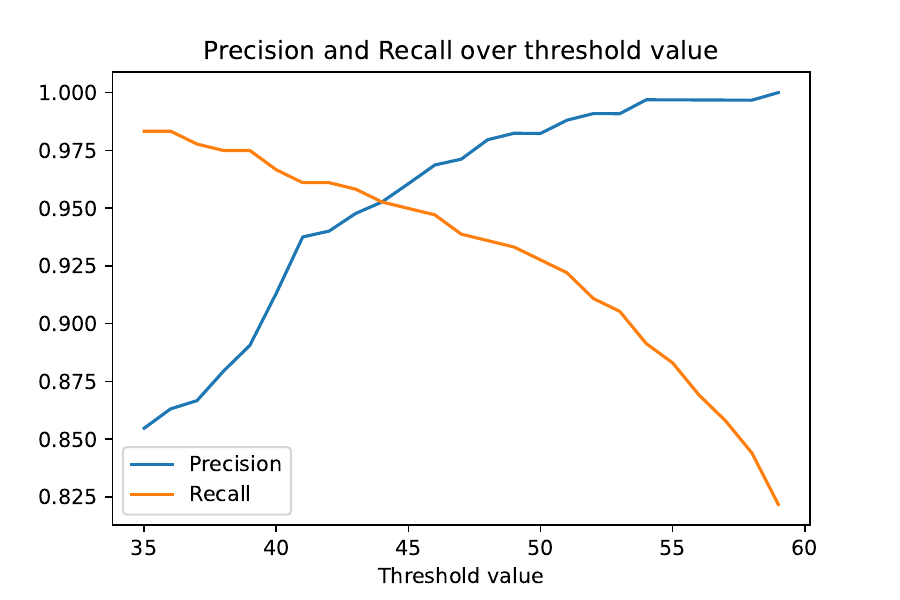}
    \caption{Precision and Recall over threshold values on the INA speaker dictionary}
    \label{fig:inadict_thresholds}
\end{figure}

Speaker identification was evaluated according to the Equal Error Rate (EER). We obtain an EER of 3.9\%\footnote{Previous work on the same corpus\cite{vallet2016speech} obtained an EER of 7.3\% using an equivalent evaluation protocol.} on the INA speaker dictionary for a similarity threshold equal to 0.40. However, because precision is more important than recall here, we decided to use a threshold equal to 0.52, at which we obtain a precision of 0.99 and a recall of 0.91 (see Figure \ref{fig:inadict_thresholds}).

\subsection{Subjective quality evaluation}
At the end of this selection process, for each target speaker, series of speech segments were available amounting to at least three minutes by speakers (i.e. for the complete corpus, that would amount to about 50 hours of speech).
In order to subjectively evaluate the quality of the automatic process, and as it is hardly feasible to listen to the complete corpus, we opted for applying a perceptual annotation on a subset of the available extracts.
The subset was composed by a random selection of one segment for each target speaker; these segments were annotated for the presence of the following potential problems: backchannel, more than one person speaking, musical background, and audible noise.
Backchannel was defined as up to two syllables produced by another speaker than the target; 
if two speakers spoke more than two syllables, it was annotated as more than one person speaking.
Audible background music or noise, while listening the extracts with headphones, were annotated as such.

The available extracts were divided in three parts, assigned to three different annotators, with a 309-large subset annotated by the three annotators.
The extracts of the common part were selected so as to propose up to 10 extracts per category of period, gender and age (let's recall we aimed at collecting 30 speakers per category); this shall amount to 320 extracts if a sufficient number of speakers were available in each category, which was not the case.
The females of 51 to 65 year-old in the 1955-1956 period, and over 65 year-old in the 1975-1976 period were only 5 and 4 in the corpus currently available, so the final number of 309 (see details in the result section for available speakers).

A python notebook was used to randomly load one extract, play it, and request the annotator to answer a series of letters indicating the presence of the potential problems. 
A field for free comments was available, before hearing the next extract.

\section{Results}

% TRAITeMENT MANUELS ont debutes a l'ina le 10 mai 2021, pause en juillet aout, deux semaines de conges en autommne, et une venue sup. en decembre:
% 22 journées à l'INA; sur lesquelles il faut décompteer un certain nnb de reuninons => si on estimee a 20 jours de W, avec 7h d'etiquettage pour A = 140h
% estimation du temps passé par rémi ? =>  lancement c'est de l'ordre de 3h par période différente (pour comprendre comment sont rangés les fichiers etc) puis 20min par nouvelle vague (une par fois où tu es venu en gros). Donc ça fait une vingtaine d'heures de lancements de scripts.

\subsection{Obtained target speakers}

Each target speaker was manually identified for one archive.
This manual identification work took about 140 hours along 20 days. 
The available speakers are detailed in table \ref{tab:spk}.
Note that it corresponds to more than 40 hours of final speech (more than 3 minutes for 874 speakers). 
A total of 533 male and 341 female targets have been identified -- thus 874 different speakers.
Sixteen categories of speakers out of 32 did not reach the aim of at least thirty speakers. 
Four groups from the 1990's and 2010's (3 females groups) were almost complete, only missing a few speakers.
For the other groups, ten have between 10 to 20 speakers (four of them are male groups), and two female groups have only 4 and 5 speakers.
A total of 211 speakers (22\% of our target) are missing, so to complete all groups with at least 30 speakers.
The sixteen groups with at least 30 speakers present 125 extra speakers.

%How much we have with comparison with the initial selection.
%Problem that lead to suppression / missing targets:
%\begin{itemize}
%    \item{target not in the archive; do not speak etc.}
%    \item{Presence of noise}
%    \item{other reasons ?  (archive not available...) }
% \end{itemize}

% Table from the perceptual evaluation, of available speakers
\begin{table}[!h]
\begin{center}
\begin{tabularx}{\columnwidth}{|X|r|r|r|r|}\hline
       %     F/M \\
       &  20-35 & 36-50 & 51-65 & $>$65  \\\hline
1955-56 &  \textbf{13}/34 & \textbf{17}/61 &  \textbf{5}/\textbf{19} & \textbf{17}/\textbf{10} \\
1975-76 &  \textbf{16}/\textbf{14} & \textbf{18}/37 & \textbf{11}/31 &  \textbf{4}/\textbf{11} \\
1995-96 &  30/\textbf{27} & 32/47 & \textbf{29}/48 & \textbf{29}/35 \\
2015-16 &  31/30 & \textbf{29}/51 & 30/48 & 30/30 \\\hline
\end{tabularx}
\caption{Number of speakers with sufficient data extracted for each period (row) and age group (columns), for gender (Female/Male). Categories with less than 30 speakers shown in boldface.}
\label{tab:spk}
\end{center}
\end{table}

For each category, more than thirty targets have been searched for, but for a series of reasons, in some cases the requirements were not met. 
These problems overwhelmingly arise for female targets, known to be under-represented in media \cite{doukhan2018describing}, and for archives from the 1950's and 1970's -- for which documentation and quality is worst.
These missing speakers are mostly linked to the following factors:
(i) the target appears in the notice, but was the topic of a program without actually appearing, or the target may not speak (or not sufficiently), or spoke in a foreign language and was interpreted.
(ii) the target may be interviewed in a noisy place (and was not detected as clean speech), or the target voice may appear during a movie trailer (and thus do not fit with our criterion of conversational speech).

A main reason more male targets were found is linked to the fact that most documents have several targets (female and male) so working on a document that features a female target generally led to the identification of one or more male targets, while the reverse is not true. 
Moreover, female targets are more prone to be presented by male speakers, without actually speaking.
A typical case is programs about cinema in the 1970's, that interviewed the director (generally a male), but just present the female actress during extracts of the movie.
For these reasons, the male categories generally present well above 30 target speakers; note that age also introduces bias in terms of presence in the media.

% \subsection{Diarization \& Speech quality estimation}
% \subsubsection{sous sous titre??}
% % includes manual selection process (not a big thing to describe)
% % must contain an estimation of the TIME it took.

% % maybe speak rapidly of the manual aspect of the process?

\subsection{Perceptual evaluation of speech quality}

% accord inter annotateur
On a subset of 309 extracts, the three annotators did evaluate the four scales. The number of extracts detected by each annotator with each type of potential problem is reported on the left part of table \ref{tab:pbs}.
The corresponding inter-annotator agreement, measured with an exact Fleiss' kappa \cite{fleiss_measuring_1971,gamer}, equals 0.629 for backchannel, 0.569 for more than one speaker, 0.855 for music, 0.448 for noise -- this amount to a kappa of 0.649 for finding a potential problem in a given extract.
These kappa values shows the relative reliability of the annotation, especially for music and backchannel.
The comparatively low kappa for noise show that noise is a more complex concept than music, and that the three annotators have somehow different views on what is a noisy extract.
% AR: je pense qu'on peut conserver seulement ceux avec *  (discuter)
%    \item Reject:       Kappa = 0.228 
%    \item Backchannel:  Kappa = 0.629 *
%    \item Speaker nb:   Kappa = 0.569 *
%    \item Gender error: Kappa = 0.569 
%    \item Music:        Kappa = 0.855 *
%    \item Noise:        Kappa = 0.448 *
%    \item Language:     Kappa = 0.776 
%    \item Other:        Kappa = 0.133 
%    \item Global:       Kappa = 0.649 

From this common ground of 309 extracts, the results on the complete set of 874 utterances was grouped.
For the utterance evaluated by three annotators, the presence of a problem was considered only if at least two of them reported it. 
The number of each error on 874 utterances are reported in the right part of table \ref{tab:pbs}.
The percentages of these four types of potential problems, on the complete corpus, and for each category of period, gender, and age, are reported in table \ref{tab:error}.

\begin{table}[!h]
\begin{center}
\begin{tabularx}{\columnwidth}{|X|r|r|r||r|}
\hline
  &  \multicolumn{3}{|l||}{Common sub-part} & Total  \\
\hline
Annotator      & A1   &    A2 &  A3 &      \\
%Reject        &   0  &    27 &  13  &   30  &(3.4) \\
\hline
Backchannel   &  57  &    82 &  44  &  148  \\
\hline
Several Spk.    &   5  &    15 &  21 &   33   \\
%Gender         &   1  &     4 &   2 &    5  &(0.6) \\
\hline
Music          &  19  &    17 &  15 &   33   \\
\hline
Noise          &  30  &    51 &  37 &   72  \\
\hline
\end{tabularx}
\caption{Number of problems, for each category, spotted by the three annotators on the 309 common sentences of the perceptual evaluation (left part), or on the complete set (right part).}
\label{tab:pbs}
\end{center}
\end{table}

The amount of observed backchannels is stable across gender and age (at about 17\%), but is higher for the 1970's, and clearly lower for the 1950's.
Presence of more than one speaker in one extract is much lower (mean below 4\%), and does not seem to be correlated to gender, while the same pattern on periods is observed (clearly higher for the 70's, lower for the 50's).
A pattern for age appears: the older the target, the less several speakers were found.

\begin{table}[!h]
\begin{center}
%\begin{tabularx}{\columnwidth}{|c|r|r|r|r|r|}
\begin{tabularx}{\columnwidth}{|X|r|r|r|r|r|}\hline
         & Bac  &  SSp  & Mus &  Noi &  Any  \\\hline
Globally & 16.9 &  3.8 &  3.8 &  8.2 &  29.7 \\\hline
1955-56  &  9.1 &  1.7 &  0.6 &  9.1 &  19.3 \\
1975-76  & 21.8 &  9.2 & 12.7 & 26.1 &  55.6 \\
1995-96  & 17.7 &  2.5 &  3.2 &  3.2 &  26.4 \\
2015-16  & 18.6 &  3.6 &  1.8 &  3.6 &  26.5 \\\hline
Female   & 18.5 &  3.5 &  4.4 & 10.0 &  32.6 \\
Male     & 15.9 &  3.9 &  3.4 &  7.1 &  28.0 \\\hline
20-35    & 17.9 &  5.1 &  2.6 &  9.2 &  29.7 \\
36-50    & 15.1 &  5.1 &  3.8 &  8.9 &  30.5 \\
51-65    & 17.2 &  2.7 &  5.4 &  6.3 &  28.5 \\
Over 65  & 18.7 &  1.2 &  3.0 &  8.4 &  30.1 \\\hline
\end{tabularx}
\caption{Percentage of utterance detected with each category of potential problem (Bac: Backchannel, SSp: Several speakers, Mus: Music, Noi: Noise, Any: presence of at least one error in the extract), globally, and for each category of period, gender, and age.}
\label{tab:error}
\end{center}
\end{table}

The presence of music varies mostly with the period, the 1970's having about 13\% of its extracts with perceivable musical background, while the other periods have lower percentages.
Only reduced changes on the presence of music are observed with age and gender.

Noise is about twice more frequent than music.
Like music, noise is particularly frequent in the 1970's (26\%) but, unlike music, is also relatively frequent in the 1950's, while low levels of noise are observed in the two more recent periods.
Slight changes of noise presence are observed across gender and age categories.

The percentage of extracts that do present at least one potential problem of 30\% globally.
This percentage varies mostly across periods, more than half of the extracts from the 1970's being annotated with potential problems, while one on five extracts from the 1950's has a potential problem.

\section{Discussion \& Conclusions}

We proposed a semi-automatic method to help selecting speakers with known characteristics (here in term of age and gender) in large media archives, avoiding silence, noise and musical background.

About 140 hours of work was invested to manually identify 915 target speakers.
Among these speakers, 41 were either not found in the documents or do not speak sufficiently to build a model of their voices (32 of these speakers had less than 20 seconds of annotated speech).
The work required for the application of the automatic processing scripts, and for processing the files (time spent by human, not by machine processing) was estimated to 20 hours for the corpus presented here.
The complete set of archives used to obtain the current set of speakers had a total duration of 453 hours.

\newcite{BAZILLON08.277} measured the time required to manually transcribe spontaneous speech at eight times the duration of the target speech; manual diarization process is more simple than a full transcription, an estimate of four times the archive duration for manual processing seems reasonable.
Due to the complexity of the target speaker identification process only (that may require an almost complete listening of the archive), this estimation does not seem unreasonable.

Given these estimations, we can assume the manual extraction of the target speech data from 453 hours of archives would have required at least more than the viewing time and up to 1800 hours of human labor.
Using the proposed method, it took about 160 hours -- i.e. four to ten times less than a manual annotation, which seems to be a fairly efficient method.
By comparison, but not on the same task, the semi-automatic transcription method proposed by \newcite{BAZILLON08.277} cut by half the manual processing time.

The perceptual evaluation of potential problems shows that one extract in three has at least one potential problem.
This may seem a relatively high rate; meanwhile, annotation of backchannel amounts for about half this number (see table \ref{tab:pbs} and shall not be a problem for the targeted analyses, as backchannels are very short regarding to the duration of extracts (about 0.1s vs. 10s for the mean duration of annotated extracts): this shall not affect the voice's spectral characteristics or mean pitch values.
Presence of music, thanks to the music detection process, was limited, compared to the frequent use of mixed music in radio and TV shows (see the evaluation part).
Moreover, even if audible wearing a headset, the levels of music that were annotated are low compared to the levels of voices.
Noise is a more complex question: we have seen its presence is difficult to annotate. 
This is certainly linked with the complexity of defining noise, that may be any audible sound added to the soundtrack but speech and music (e.g. street noise, steps, natural noises), but also sounds linked to the recording place (echo from the room), from the recording equipment, from the many hardware used to archive the media (disk, tape), or from compression used to store audio files.
The fact noises are more present in the 1970 cue for high presence of added noise, because the selected programs happen to have more outdoor recordings, for example (observation made during the manual target spotting).
On the contrary, the lower levels of noise in the 1950's compared to the 70's (somehow counterintuitive) shows archive processing at INA shall not introduce major bias -- even if more recent media have better quality.

Noise and music potential effects on the targeted measures will be evaluated once the corpus is completed, but this is outside the scope of this paper.
The presented methodology may apply for the construction of corpora dedicated to e.g., sociological work, that do not necessarily require high sound quality.

Presence of several speakers in one extract is an unwanted feature, and ideally these extracts shall be removed.
It'll be important to screen the extracts labelled as such to estimate the relative ratio of extracts with completely different speakers, compared to extracts featuring speakers with comparable voices.
The second case is less problematic than the former.
During informal testing to set up the perceptual evaluation, one extract was labelled as featuring two different speakers only by one of the three annotators: the other two hadn't noticed -- only the dialogue's semantics allowed judging there were two speakers.
The fact the kappa for this measure was 0.57, shows in a good deal of the cases, the difference in voices was not spotted by all annotators, which pleads for similarity between the voices. 
The evaluation of the speaker identification system, with higher similarity within than across genders shows it is a probable outcome.

Identification work will continue until having a complete set of speakers.
Then, evaluations of the output quality will be applied, with estimation of signal-to-noise ratios, potential distortion of acoustic measurements due to music background, etc. 
This corpus has a vocation to be shared via INA's online resource management system\footnote{\url{https://dataset.ina.fr/}}, once the project will be over.
The software developed to apply the processing described here shall also be made public in the future, after consolidation of their use on other corpus building.

\section{Acknowledgements}

This work has been partially funded by the French National Research Agency (project Gender Equality Monitor - ANR-19-CE38-0012).

% \nocite{*}
\section{Bibliographical References}\label{reference}
%\label{main:ref}

\bibliographystyle{lrec2022-bib}
\bibliography{lrec2022-CorpusGenre}

\section{Language Resource References}
\label{lr:ref}
\bibliographystylelanguageresource{lrec2022-bib}
\bibliographylanguageresource{languageresource}

\end{document}